\documentclass{elsart}
\usepackage{graphicx}
\usepackage{amssymb}
\usepackage{amsmath}
\begin{document}
\bibliographystyle{elsart-num}
\begin{frontmatter}
\title{\vspace*{-2cm} {\small {\rm BMC Bioinformatics} {\bf 4}{\rm , 580 (2003)}}\\ \vspace*{1cm}
Online tool for the discrimination of equi-distributions}
\author{Thorsten P\"oschel$^*$, Cornelius Fr\"ommel, and Christoph Gille}
\address{Charit\'e, Institut f\"ur Biochemie,\\
  Monbijoustra{\ss}e 2, D-10117,  Berlin, Germany\\thorsten.poeschel@charite.de, 
  cornelius.froemmel@charite.de, christoph.gille@charite.de\\
$^*$corresponding author\\ \today}
\begin{abstract}
{\bf Background:}  For many applications one wishes to decide whether a certain set of
  numbers originates from an equiprobability distribution or whether
  they are unequally distributed. Distributions of relative
  frequencies may deviate significantly from the corresponding probability
  distributions due to finite sample effects. Hence, it is not trivial
  to discriminate between an equiprobability distribution and
  non-equally distributed probabilities when knowing only frequencies.\\
\noindent {\bf Results:} Based on analytical results we provide a software tool which allows
  to decide whether data correspond to an equiprobability
  distribution. The tool is available at http:/$\!$/bioinf.charite.de/equifreq/.\\
\noindent {\bf Conclusions:} Its application is demonstrated for the distribution of point mutations in coding genes.
\end{abstract}

\end{frontmatter}

\section*{Background}

\begin{figure}
\centerline{\includegraphics[width=8cm,clip]{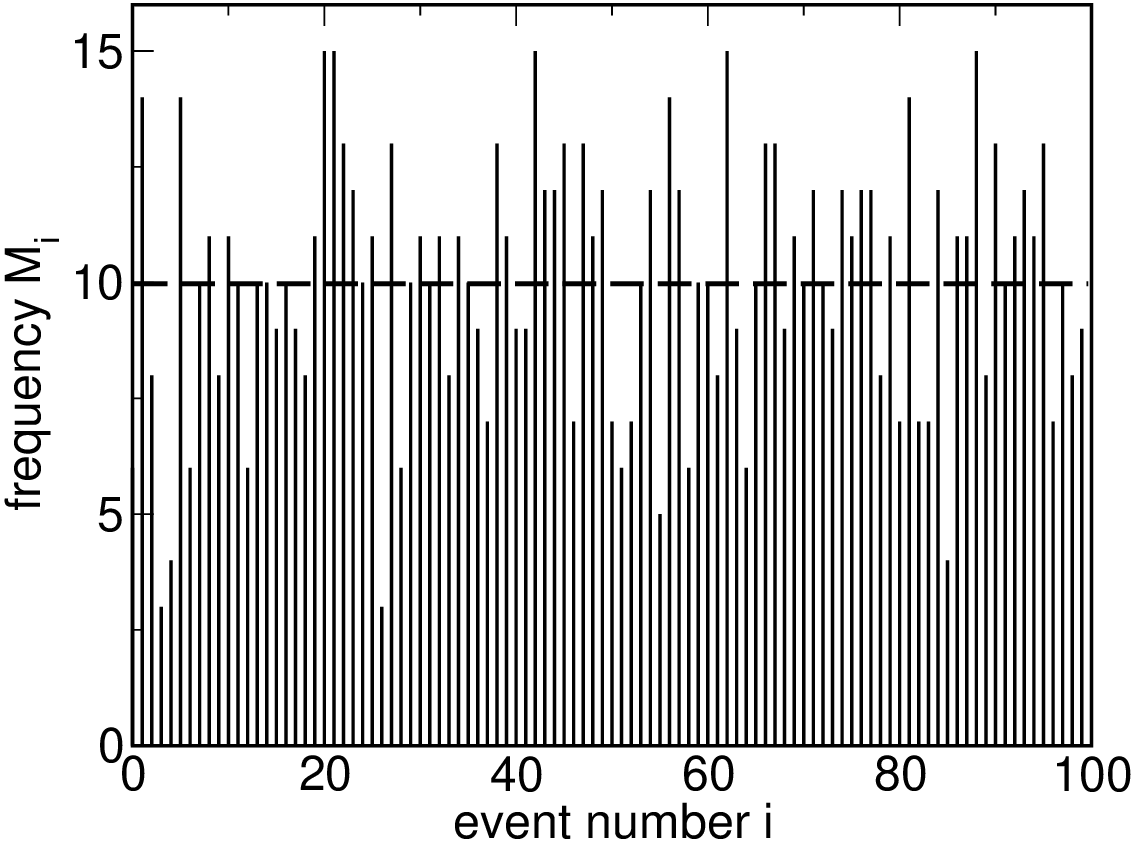}}
\caption{Histogram of frequencies of $M=1000$ events which have been drawn according to an equiprobability distribution $p_i=1/N=1/100$. The dashed line displays the expectation value.}
\label{fig:freqs}
\end{figure}

Assume a set of certain events occur with frequencies $M_i$, $i=1\dots
N$, with $\sum\limits_{i=1}^N M_i=M$, e.g.,
$M_i=\{4,5,2,3,2,9,3,3,5,12,4,6,4,\dots\}$. We ask the question
whether the events obey an equiprobability distribution $p_i\equiv
1/N$. According to the general definition of probabilities
\begin{equation}
p_i=\lim\limits_{M\rightarrow\infty}\frac{M_i}{M}\,,
\end{equation}
for an equiprobability distribution and for large sample size $M$ it
is expected to find each of the events approximately $M_i\equiv M/N$
times. For finite sample size, however, the frequencies $M_i$ may
deviate considerably from this value (Fig. \ref{fig:freqs}).

\begin{figure}
\centerline{\includegraphics[width=8cm,clip]{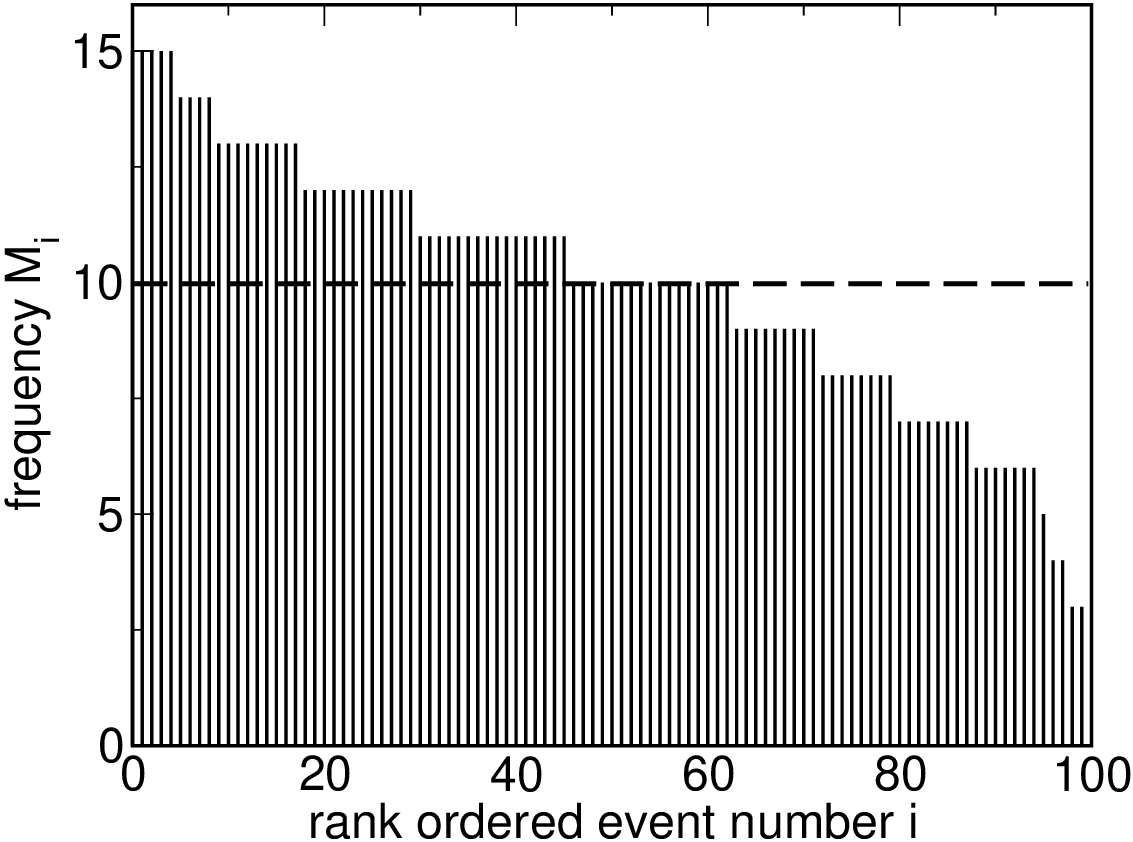}}
\caption{Same data as in Figure \ref{fig:freqs} but in rank order. 
  From the figure it might be erroneously concluded that the events do
  not obey an equidistribution. The distribution is deformed, however,
  exclusively due to finite sample effects.}
\label{fig:freqsordered}
\end{figure}

The deviation from the equidistribution becomes particularly obvious
if we order the events according to their rank, i.e., the most
frequently occurring event appears left at the abscissa, then the next
frequent, etc. (Fig. \ref{fig:freqsordered}).

If we conclude na\^{\i}vely from the observed frequencies to the
probabilities, i.e., if we assume $p_i/p_j=M_i/M_j$, in the extreme
case $M_{100}=3$ we end up with a relative error of 70\,\%.  In other
words, from the frequencies measured in an experiment as shown in
Figs. \ref{fig:freqs} and \ref{fig:freqsordered}, it might be
erroneously concluded that the events are strongly non-equally
distributed.

Using the methods of statistics we can generate (predict) the rank
ordered frequency distribution for given $N$ and $M$ under the
precondition that the events are equidistributed \cite{TPJanf}. The
predicted frequency distribution can then be compared with the
distribution as measured in an experiment with the same values of $M$
and $N$. From the comparison it can be judged whether the events in
the experiment obey an equidistribution.

Following this procedure we describe a tool which helps to decide
whether a given set of frequencies complies with an equidistribution.
For demonstration the tool is applied to the distribution of point 
mutations in human genes.

\section*{Implementation}

The numerical tool is available via the web address \\
http:/$\!$/bioinf.charite.de/equifreq/. The underlying kernel program which computes the most probable frequency distribution is implemented in $C++$ and the user interface is written in PHP. The program source  is available at  this address.

\section*{Results and Discussion}
\subsection*{Mathematical method}
We want to sketch briefly the derivation of the basic formula: Assume
we distribute $M$ balls over $N$ urns according to an
equidistribution. The probability $p(k_i,i)$ to find $k_i$ urns filled
each with {\rm exactly} $i$ balls is given by
 \begin{equation}
p(k_i,i) = \frac{M!}{N^M}\sum\limits_{j=k_i}^{\lfloor M/i\rfloor}
(-1)^{(j-k_i)}
\binom{j}{k_i}\frac{(N-j)^{(M-ji)}}{(i!)^{j} (M-ji)!}\,,
\label{eq:InclExcl}
\end{equation}
where $\lfloor x\rfloor$ denotes the integer of $x$.

Note that the probability to find a number of $k_i$ urns which contain
{\em exactly} $i$ balls is different form the probability to find the
number of urns which contain {\em at least} $i$ balls which is a
simple textbook problem, whereas the derivation of Eq.
\eqref{eq:InclExcl} requires quite involved algebra. The relation
between both probabilities is provided by the exclusion-inclusion
principle \cite{vonMises,JK}. For our purpose we need the number
$\left<K_i\right>$ of urns filled with $i$ balls which are found on
average, i.e., we need the first moments of the probabilities Eq.
\eqref{eq:InclExcl}. These values can be found in closed form applying
the method of generating functions for the descending factorial
moments. The averages $\left<K_i\right>$ have been derived in a
different context earlier, the details of the derivation can be found
in \cite{JanfTP,TPJanf}:
\begin{equation}
 \langle K_i\rangle= 
 N\binom{M}{i} \frac{1}{N^i} \left(1-\frac{1}{N}\right)^{(M-i)}\,.
\label{eq:Mom}
\end{equation}
As an interesting detail of the solution, the average number of filled
urns is given by the total number of urns minus the number of empty
ones, $N^*=N-\left<K_0\right>$, i.e. \cite{TPJanf},
\begin{equation}
\label{eq:Nstar}
\frac{N^*}{N}
=1-\left(1-\frac{1}{N}\right)^M
\approx 1-\exp\left(-\frac{M}{N}\right)\,.
\end{equation}
Obviously, for small $M$ (numbers of balls) there is a significant
number of urns which, on average, stay empty. Translating back to the
language of biology we come to a surprising result: given a population
of $N=1000$ species. If we investigate a number of $M=5000$
individuals, from Eq. \eqref{eq:Nstar} we obtain $N^*\approx 993.3$,
i.e., about 7 species are never found, although from na\^{\i}ve
reasoning one expects each species occurring about 5 times.

The moments $\left<K_i\right>$ given in Eq. \eqref{eq:Mom} allow to
reconstruct the rank ordered frequency distribution since they
describe how many, on average, events do not occur (zero times), how
many occur once, twice, etc. Hence, the desired rank ordered frequency
distribution reads finally
\begin{equation}
M_i^{\rm theo}=
\begin{cases}
0~~ \text{for} & N\ge i > N-\left<K_0\right>\\
1~~ \text{for} & N-\left<K_0\right> \ge i  >  N-\left<K_0\right> -\left<K_1\right> \\
&\dots\\
j~~ \text{for} & N-\sum\limits_{k=0}^{j-1} \left<K_k\right> \ge i 
 > N-\sum\limits_{k=0}^{j}\left<K_k\right>\,.
\end{cases}
\label{eq:Hauf}
\end{equation}

\begin{figure}
\centerline{\includegraphics[width=8cm,clip]{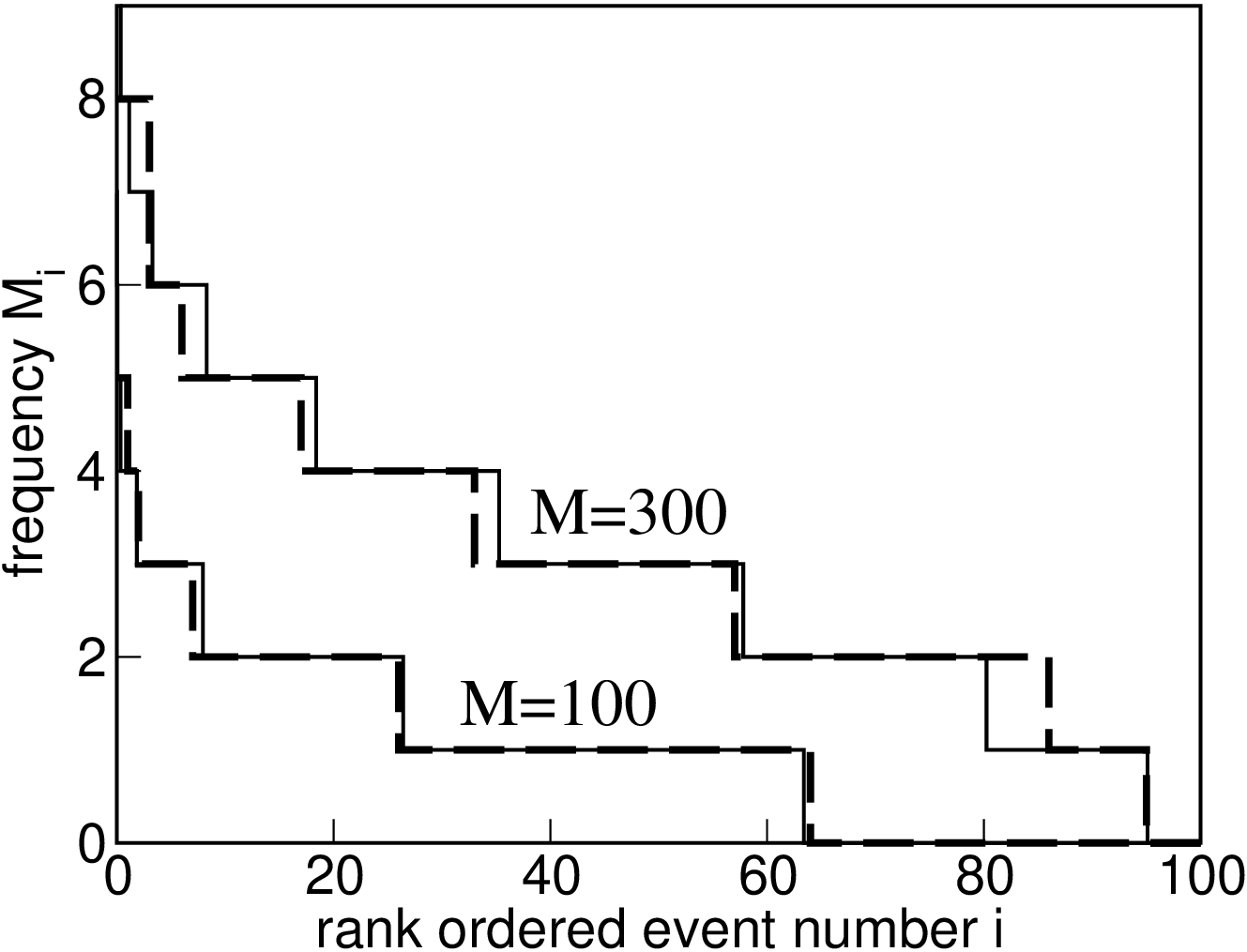}}
\caption{Rank ordered frequency distribution for $N=100$ equally 
  distributed events for different sample size $M$. The solid lines
  show the distribution as predicted by Eq. \eqref{eq:Hauf}, the
  dashed lines show the distribution of independently drawn
  equidistributed random numbers from the interval $[1,N]$.}
\label{fig:prove}
\end{figure}

\begin{figure}
\centerline{\includegraphics[width=8cm,clip]{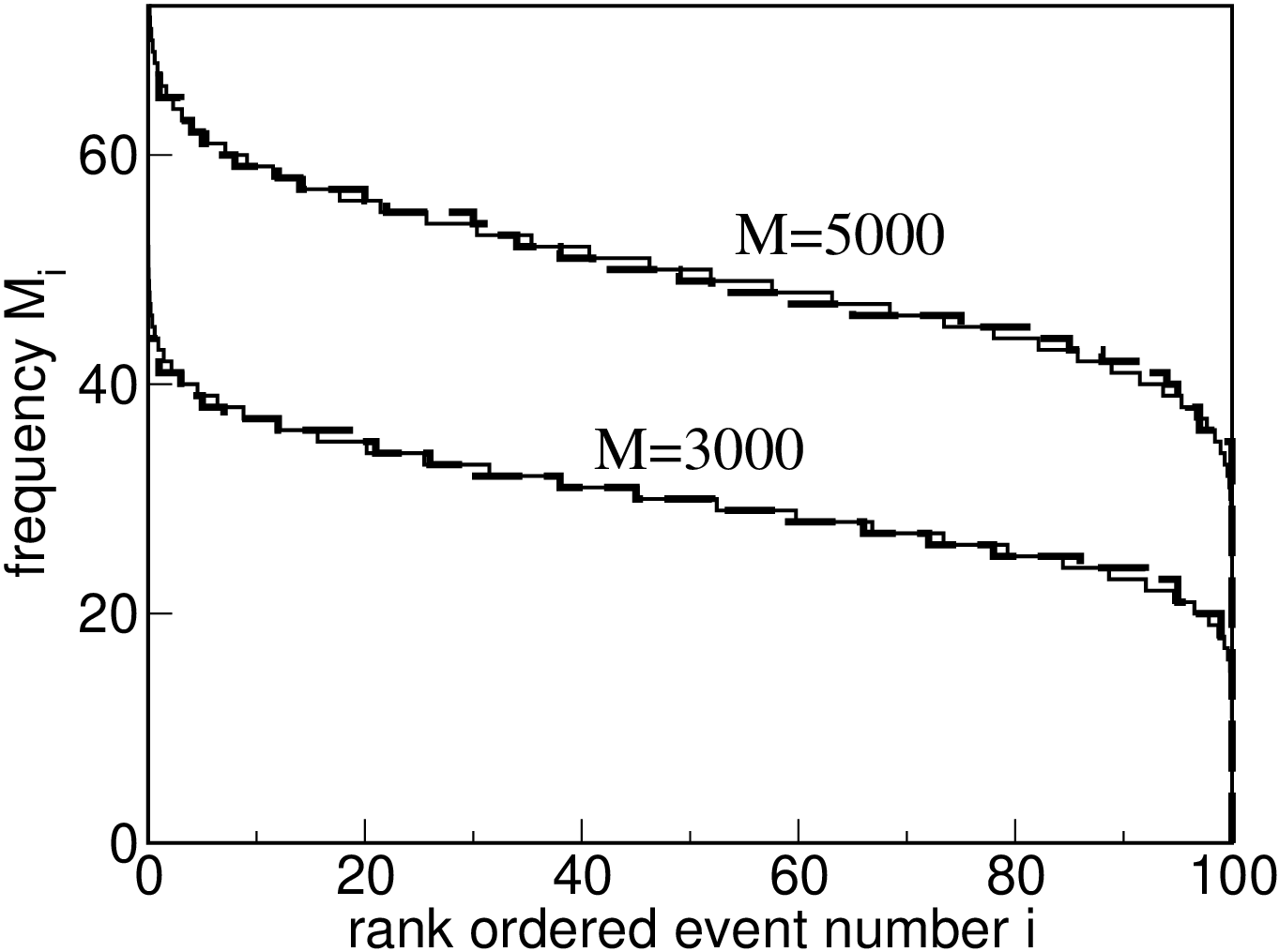}}
\caption{Same as Fig. \ref{fig:prove} but for larger sample size $M$.}
\label{fig:prove1}
\end{figure}

We apply Eq. \eqref{eq:Hauf} to predict the frequency distribution
which arises from an equidistribution for different sample sizes $M$
and compare with direct numerical simulations, s. Figs.
\ref{fig:prove}, \ref{fig:prove1}. The predictions due to Eq. \eqref{eq:Hauf} agrees
well with the numerical experiment.

\subsection*{Exploration of experimental data}

The theoretical distribution of frequencies due to Eq. \eqref{eq:Hauf}
can be compared with experimentally obtained frequencies. From the
distance between both (rank ordered) frequency distributions we can
conclude whether the experimental data obey an equidistribution. To
this end we have elaborated a web based tool (http:/$\!$/bioinf.charite.de/equifreq/).
The user
interface offers four alternative input masks which differ in the way
the input file is generated:
\begin{enumerate}
\item The measured frequencies of each species $M_i$ are given directly.
\item The number of species $N$ and the total number of individuals
  $M$ are specified. Each individual is assigned a species by chance.
\item As for (2) the rank ordered frequencies are computed but with
  the generalization that each species is assigned an individual
  probability. The theoretical basis for this computation is not given
  here but will be published elsewhere \cite{biofreq}.
\item The last input mask is intended for the investigation of the
  spatial distribution of point mutation in genes which is presently
  the most specialized application of the described program.
\end{enumerate}

The program computes  the expected frequency
  distribution due to Eq. \eqref{eq:Hauf} with the assumption that the
  species obey an equiprobability distribution.
Three output files are generated: {\em freq}, {\em ktheo} and
{\em kexp}. The file {\em freq} contains the rank ordered frequencies
as generated from the input data set (cases (1) and (4)) or randomly
due to an equiprobability distribution (case (2)) or a general
distribution (case (3)). {\em ktheo} contains the moments
$\left<K_i\right>$ for each rank $i$, i.e. the expected number of
individuals occurring $i$ times, due to Eq. \eqref{eq:Mom} for given
numbers $N$ of species and $M$ of individuals. For cases (1) and (4)
the values of $M$ and $N$ are extracted from the input data, for (2)
and (3) they are provided by the user.  (Note that these expectation
values are real numbers in general.) The third column of line $i$
contains the value $M-\sum_{j=0}^i\left<K_i\right>$. The last file,
{\em kexp} contains the same data as {\em ktheo}, but based on the
input data (cases (1) and (4)) or on the randomly generated data
(cases (2) and (3)), respectively. Besides the pure output files the
program generates a number of visualizations 
(see section {\em Example: Distribution of point mutation in genes}).  
In order to compare the experimental data with
the mathematical prediction both, the experimental data and the
theoretical data, are plotted in the same chart. Congruence of both
curves indicates that the experimental data obey an equidistribution
(case (2)) or the specified distribution (case (3)), respectively.

It may occur that the curve of the rank ordered experimental data
decays significantly slower than the corresponding theoretical curve due
to Eq. \eqref{eq:Hauf}. Since there is no distribution more
homogeneous than the equidistribution this situation may occur either
as a rare fluctuation (recall that the theoretical curve was generated
according to the {\em averaged} occupation numbers, Eq.
\eqref{eq:Mom}).  In such cases there is no probability distribution
$\{p_i\}$ which reproduces the experiment {\em on average}. This case
can be artificially evoked when the species in the input file occur
with almost identical frequencies.

The difference between the experimental rank ordered frequency
distribution and the corresponding theoretical distribution (Eq.
\eqref{eq:Hauf}) evaluates the degree of coincidence of the input data
with an equidistribution (case (1)) or with a specified distribution
(case (3)). We define the score by
\begin{equation}
  \label{eq:score}
  S=\sum_{i=0}^M \left|M_i-M_i^{\rm theo}\right|\,.
\end{equation}
The significance of a particular difference score can be assessed by
relating it to the distribution of difference scores. This
distribution depends on $M$ and $N$. 

\subsection*{Example: Distribution of point mutations in genes}

The increasing number of known point mutations and polymorphisms in many genes coding
for pathogenetically important proteins offers the opportunity to
apply statistical tests to correlate their type and location to
evolutionary, biological and clinical features.

In each replication generation there occur mutations of the genome but 
frequently they remain unnoticed since they do not cause diseases. 
These so-called polymorphisms or variants may occur either in regions of the 
genome which are coding for amino acid sequences or in non-coding segments. 
Those changes of the DNA sequence that alter the amino acid sequence are 
frequently associated with diseases because the respective proteins cannot 
operate properly.  
Screenings for mutations using DNA of patients 
have been performed for many human diseases and the identified mutations  are 
accessible in mutation databases \cite{database}.

The detection of so-called mutation hot spots, i.e. sequence regions
with many mutation positions, is important for the identification of
the functional and genetical properties of the genetic code
\cite{function}. These hot spots must be distinguished from statistical
fluctuations that occur even when the probabilities for mutations are
identical for each residue position. Moreover, the spatial distribution
of point mutations in genes is of importance for the localization of
coding and non-coding parts in the genome.

We wish to apply the described method to the investigation of the
amino acid sequence of the cystic fibrosis transmembrane conductance regulator.
The unperturbed gene
(wild type) is given as a sequence of 1480 letters: {\small
  MQRSPLEKASVVSKLFFSWTRPILRKGYRQRLELSDIYQIPSVDSADNLSEKLER\dots}, each
standing for one amino acid \cite{cysticsequence}. In experiments
there has been observed a large number of mutations, i.e., deviations
from this sequence. Such mutations are available from  data
bases, e.g.  \cite{database}.

$\begin{array}[b]{c} M1V \\ M1K \\ \underbrace{M1I }{} \\ {\bf M} \end{array} $
$\begin{array}[b]{c} \underbrace{Q2X }{} \\ {\bf Q} \end{array} $
${\bf R}$ 
$\begin{array}[b]{c} \underbrace{S4X }{} \\ {\bf S} \end{array} $
$\begin{array}[b]{c} \underbrace{P5L }{} \\ {\bf P} \end{array} $
${\bf L}$ 
${\bf E}$ 
${\bf K}$ 
${\bf A}$ 
$\begin{array}[b]{c} \underbrace{S10R }{} \\ {\bf S} \end{array} $
${\bf V}$ 
${\bf V}$ 
$\begin{array}[b]{c} \underbrace{S13F }{} \\ {\bf S} \end{array} $
$\begin{array}[b]{c} \underbrace{K14X }{} \\ {\bf K} \end{array} $
${\bf L}$ 
${\bf F}$ 
${\bf F}$ 
${\bf S}$ 
$\begin{array}[b]{c} W19C \\ \underbrace{W19X }{} \\ {\bf W} \end{array} $
${\bf T}$ 
${\bf R}$ 
${\bf P}$ 
${\bf I}$ 
${\bf L}$ 
${\bf R}$ 
${\bf K}$ 
$\begin{array}[b]{c} G27X \\ \underbrace{G27E }{} \\ {\bf G} \end{array} $
$Y$ 
$R$ 
$\begin{array}[b]{c} \underbrace{Q30X }{} \\ {\bf Q} \end{array} $
$\begin{array}[b]{c} R31C \\ \underbrace{R31L }{} \\ {\bf R} \end{array} $
${\bf L}$ 
${\bf E}$ 
${\bf L}$ 
${\bf S}$ 
${\bf D}$ 
${\bf I}$ 
${\bf Y}$ 
$\begin{array}[b]{c} \underbrace{Q39X }{} \\ {\bf Q} \end{array} $
${\bf I}$ 
${\bf P}$ 
$\begin{array}[b]{c} \underbrace{S42F }{} \\ {\bf S} \end{array} $
${\bf V}$ 
$\begin{array}[b]{c} \underbrace{D44G }{} \\ {\bf D} \end{array} $
${\bf S}$ 
$\begin{array}[b]{c} \underbrace{A46D }{} \\ {\bf A} \end{array} $
${\bf D}$ 
${\bf N}$ 
${\bf L}$ 
$\begin{array}[b]{c} S50P \\ \underbrace{S50Y }{} \\ {\bf S} \end{array} $
${\bf E}$ 
${\bf K}$ 
${\bf L}$ 
${\bf E}$ 
${\bf R}$\dots 

The codes on top of the underbraces stand for the found mutations,
e.g., $P5L$ means that at position 5 it has been found that the
amino acid proline (P) was replaced by leucine (L).

\begin{figure}[htbp]
\centerline{\includegraphics[width=8cm,clip,angle=0]{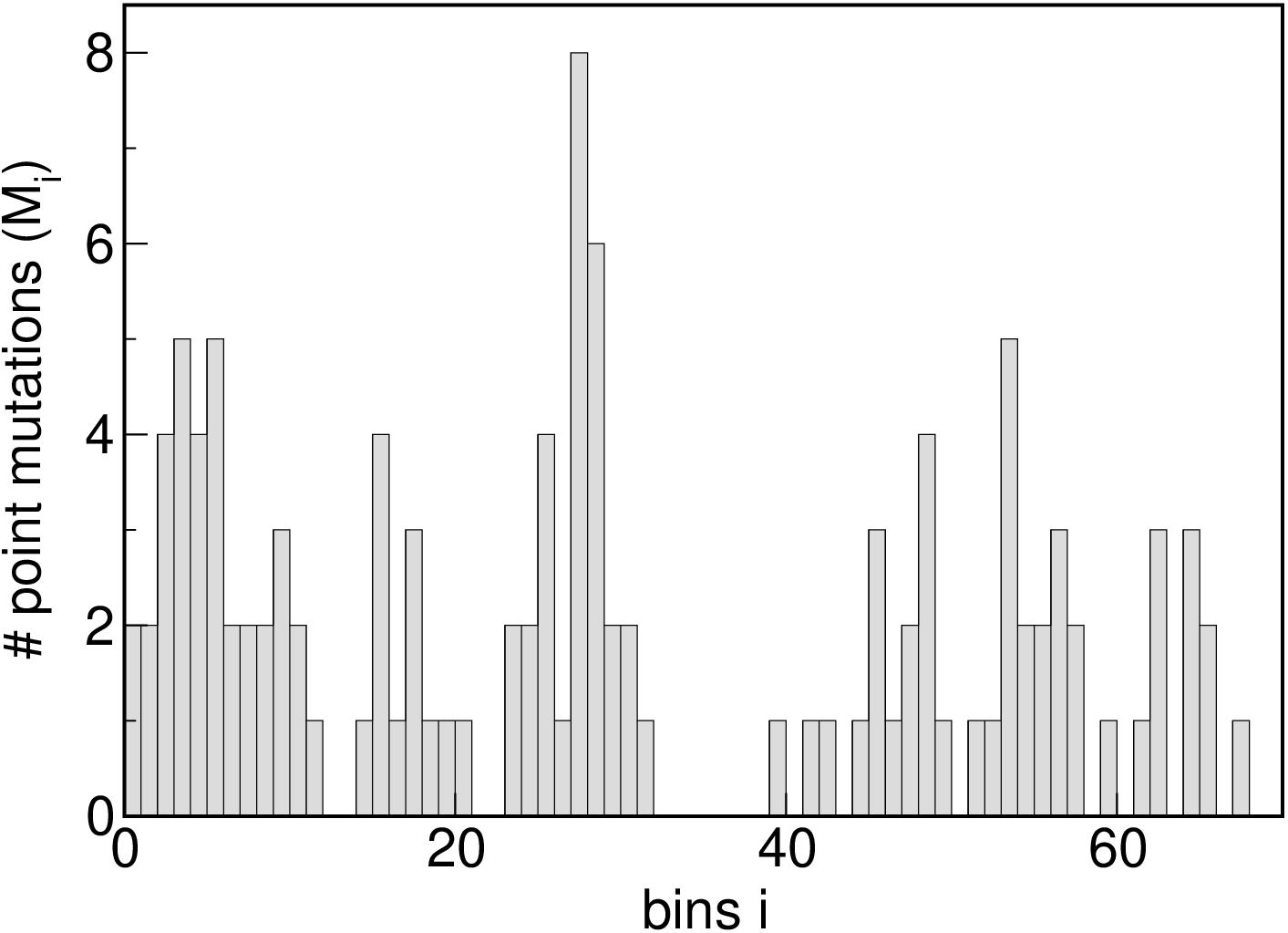}}
  \caption{Observed numbers of point mutations. The sequence has been 
    subdivided into 74 parts of length 20.}
  \label{fig:CFroh}
\end{figure}

We subdivided the sequence into 74 parts of equal length 20 and
counted the number of point mutations in each part. This way we obtain
the {\em measured frequencies}
$M_i=\{2,2,4,5,4,5,2,2,2,3,2,1,0,0,1,4,\dots\}$ which serve as input
data. (The subdivision into parts may be repeated with a different
starting point which yields similar results.) Certainly, measured
frequencies as small as given above do not allow for the application
of the $\chi^2$-test. The measured frequencies are shown in Fig.
\ref{fig:CFroh}.
Obviously, based on this data it is not possible to decide \`a priori
whether the frequencies are equidistributed.

\begin{figure}[htbp]
\centerline{\includegraphics[width=8cm,clip,angle=0]{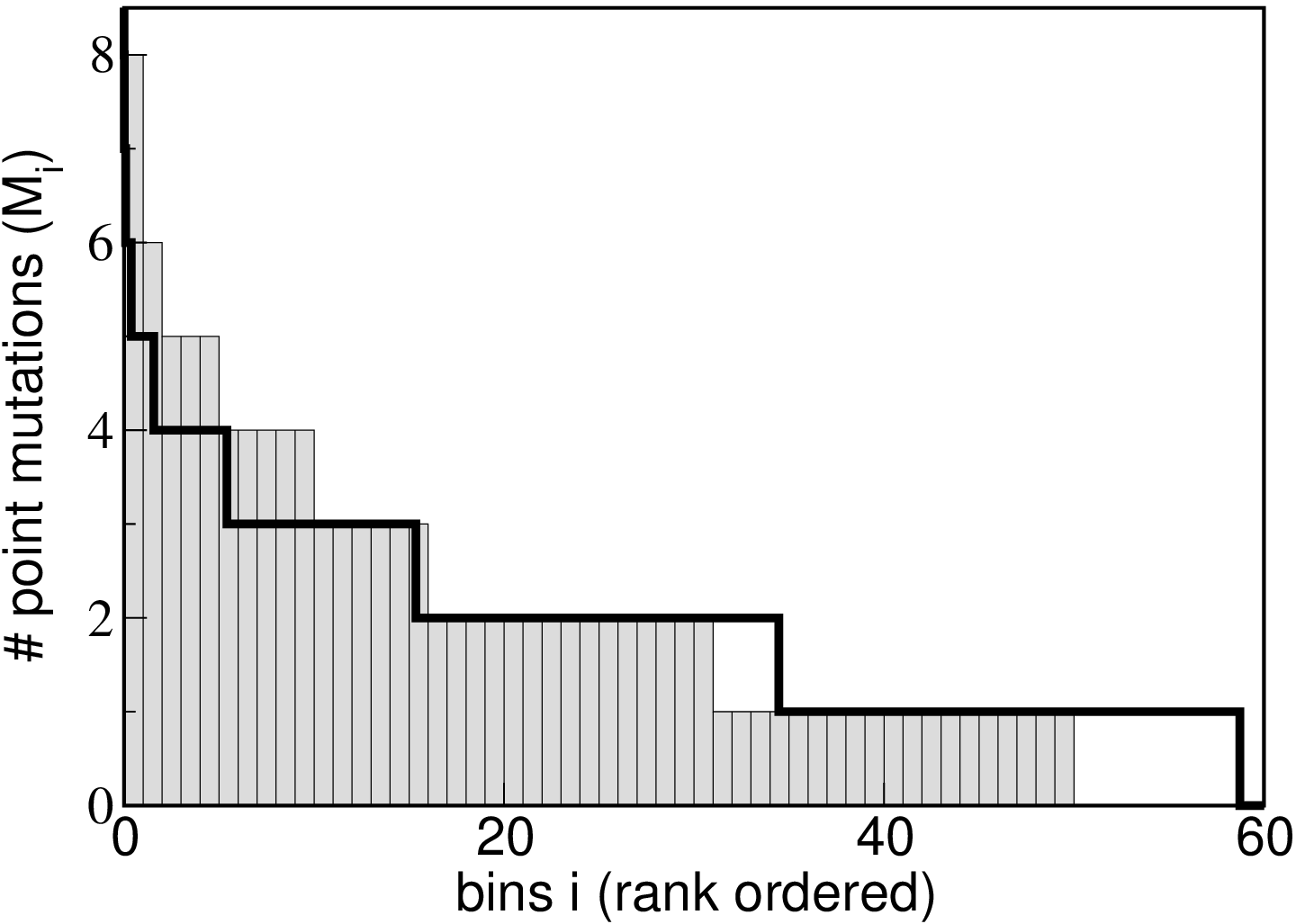}}
  \caption{Results of the computation. The bars show the rank 
    ordered frequencies, the line displays the expected frequency
    distribution which would be obtained if the point mutations were
    (equally) randomly distributed. The curves differ significantly,
    therefore, we conclude that the point mutations are not random.}
  \label{fig:CFresult}
\end{figure}

After processing the data as described above we obtain the rank
ordered measured distribution (bars in Fig. \ref{fig:CFresult}).
The full line shows the expected (theoretical) frequency distribution
due to Eq. \eqref{eq:Hauf} which has been generated with the
hypothesis that the positions of the point mutations are
equidistributed. Both curves deviate significantly from each other,
therefore, we conclude that the mutations are not equidistributed.
This conclusion agrees with the hypothesis in ref. \cite{cystichypothesis}.

Since the investigation of point mutation is an interesting field
of application of the program we developed a separate input mask for
this purpose (case (4) of the list in the previous section). The input
syntax for this mode is described in detail in the online help file of the
program.

\begin{figure}[h]
\centerline{\includegraphics[width=8cm,clip,angle=0]{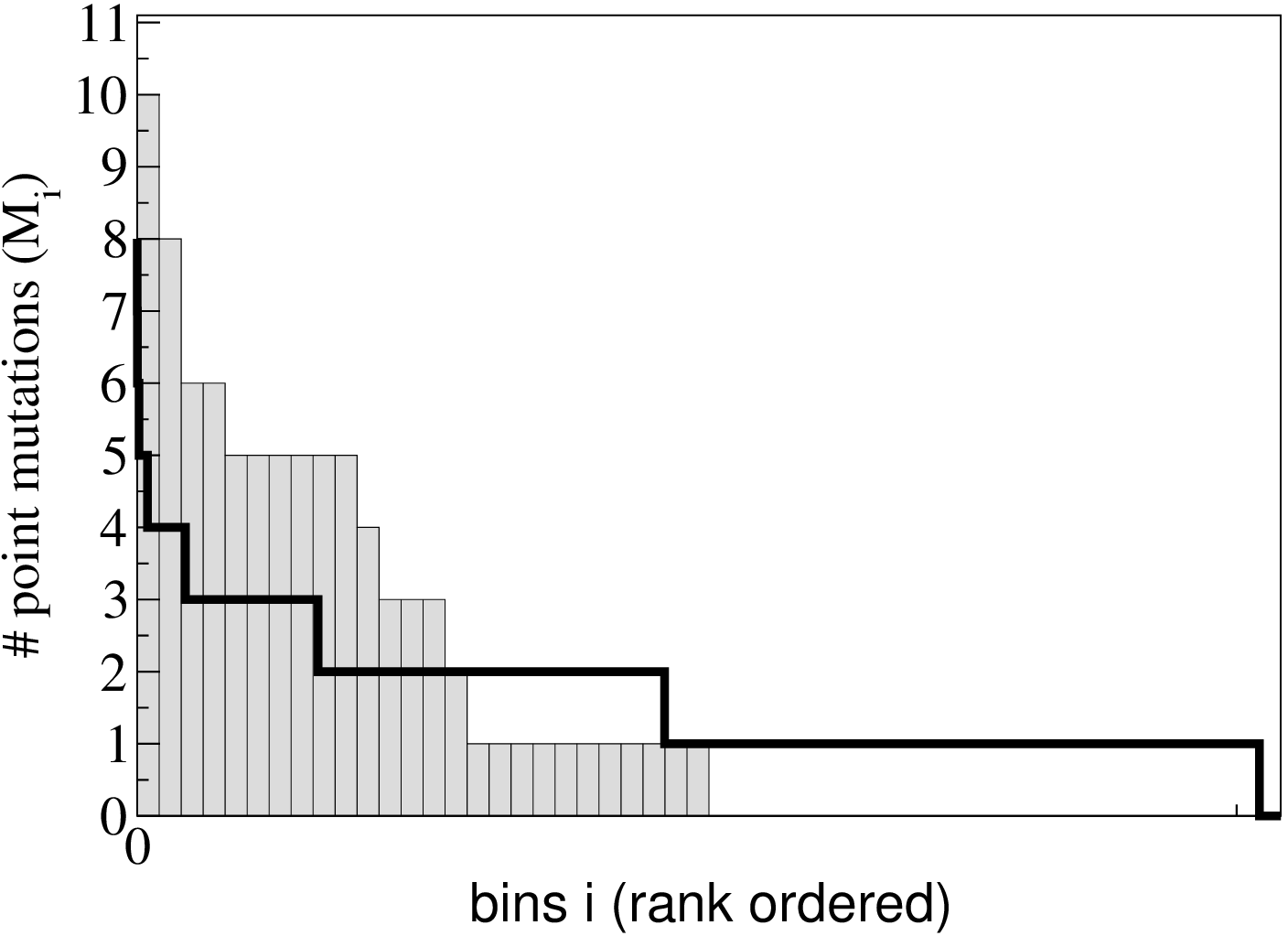}}
  \caption{Distribution of missense mutations in the androgen receptor for
    germline mutations leading to AIS. The height of the bars reflects the rank
    ordered frequencies of mutations in sequence intervals of length 20. The thick line
    displays the expected frequencies which would be obtained if point
    mutations were randomly distributed.}
  \label{fig:androgen}
\end{figure}

\begin{figure}[h]
\centerline{\includegraphics[width=8cm,clip,angle=0]{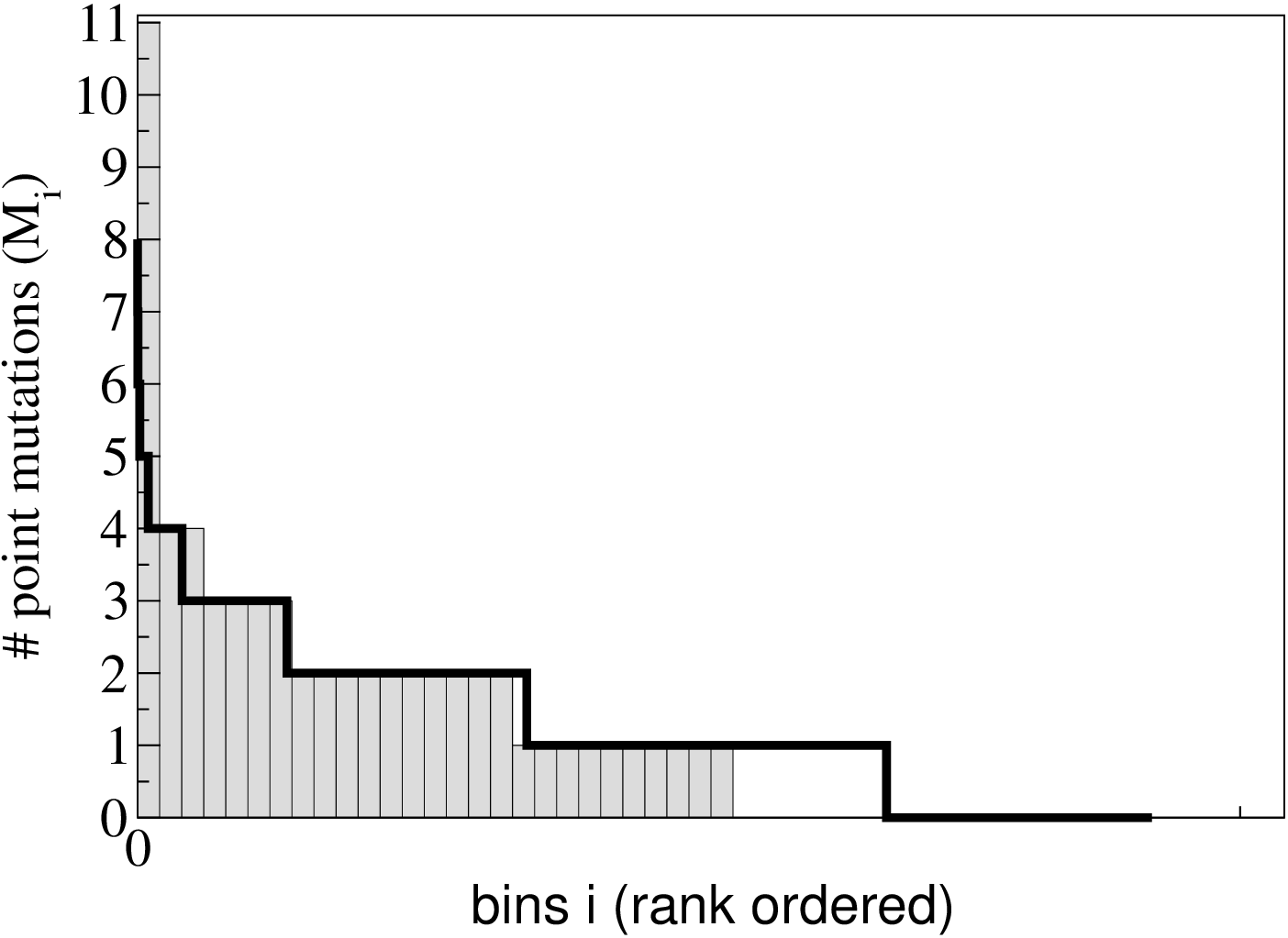}}
  \caption{Distribution of missense mutations in the androgen receptor for somatic mutations
    associated with prostate cancer. Explanation see caption of Fig. \ref{fig:androgen}.}
  \label{fig:androgen1}
\end{figure}

Recently, it has been shown for point mutations 
in the human androgen receptor (AR) that  the severity of the disease correlates with the local 
sequence conservation  \cite{androgen}. 
Germline mutations in the gene of the androgen receptor lead to the 
androgen insensitivity syndrome (AIS). 
%
%
%
%
%
In addition it was found that somatic point mutations associated with
prostate cancer are more frequently found at locations with higher
sequence variation compared to germline mutations leading to complete
AIS.
The related prediction method SIFT \cite{sift} has been proposed
recently. Both methods, SIFT and the method used in \cite{androgen}
are based on the alignment of a large number of related proteins.
Inspired by their observation we asked the question whether mutations
in the androgen receptor are distributed randomly over the sequence
depending on the association with AIS or prostate cancer.
The disease-associated mutations in the AR were obtained from the AR
gene mutation database \cite{AndrogenDatabase}.
Multiple mutations  at identical positions were counted only once.
Those mutations  resulting in  single amino acid substitutions
were included in the analysis.
The test was performed for 61 mutations associated with prostate
cancer and 86 mutations found in patients with complete AIS.
To perform the  analysis we divided the sequence of 919 amino
acids into 46 intervals of length 20 and counted the number of
mutations in each interval.
As expected, the results for the two datasets were
different:
Cancer associated
mutations are more disseminated than congenital mutations found in
patients with AIS.   
For  mutations associated with prostate cancer the bar chart of the rank
ordered frequencies nearly follows the theoretical curve for equal
probabilities (Figs. \ref{fig:androgen}, \ref{fig:androgen1}) whereas for AIS associated
mutations the bar chart deviates markedly from the theoretical curve.
Based on this finding we hypothesize that  mutagenesis in the germline is
followed by a selection process so that only a portion of the mutations are
found in patients  while others lead to early embryonal or fetal death.
Conversely, mutations associated with prostate cancer may persist and are recorded.

\section*{Conclusions}

For small sample sizes the relative frequencies $M_i/M$ of occurrence
of individuals of a certain species $i$ deviate significantly from the
probabilities of occurrence $p_i$. With the assumption that the $N$
species occur with equal probability $p_i=1/N$ the expectation values
$\left<K_j\right>$ of the numbers of events which are contained $j$
times ($j=0,\dots,M$) in a sample of $M$ individuals can be determined
based on combinatorial algebra. These expectation values allow for a
prediction of the rank ordered frequency distribution.

For many practical problems the amount of available data is
insufficient to employ standard tests, such as $\chi^2$, to
discriminate whether or not a certain set of events complies with an
equiprobability distribution. For such situations which occur
frequently in the biological sciences we have developed an online tool
which is available at http:/$\!$/bioinf.charite.de/equifreq/.
As demonstrated for the case of
point mutations in the sequence of amino acids of 
the cystic fibrosis transmembrane conductance regulator and the androgen receptor,
even for sample set sizes which are certainly not sufficient to decide
this question directly from the observed frequencies (see Figs.
\ref{fig:prove}, \ref{fig:prove1}) this tool helps to make a reliable statement.

The proposed method may be generalized to arbitrary probability
distributions provided there exists a hypothesis on the functional
form of the distribution \cite{PER}. For mathematical reasons,
however, (see \cite{biofreq}) it is more difficult to derive an
equivalent to Eq. \eqref{eq:Hauf} formula for non-equiprobability
distributions, which is subject of current research.

\section*{Avalability and requirements}
\begin{itemize}
\item {\bf Project name:} equifreq
\item {\bf Project home page:}  http:/$\!$/bioinf.charite.de/equifreq/
\item {\bf Operating systems:} platform independent
\item {\bf Programming language:} C++
\item {\bf Other requirements:} none
\item {\bf License:} GNU GPL
\item {\bf Any restrictions to use by non-academics:} none
\end{itemize}

\section*{Authors' contributions}
TP worked out the statistical and combinatorial background, wrote the kernel C++-program and drafted the manuscript. CF and CG provided the biological expertise, collected relevant biological data and organized the biological relevant applications. CG wrote the PHP user interface.
All authors contributed in writing the manuscript.

\section*{Acknowledgments}
The authors are grateful to W. Ebeling, J. Freund and R. Mrowka for
helpful discussion.
We thank the reviewers for their helpful remarks and   recommendations.
Particularly fruitful was the analysis of  mutations in the androgen receptor.

\end{document}